\documentclass{article}
\usepackage[affil-it]{authblk}
\usepackage{amsmath, amsfonts, amsthm, amssymb}
\usepackage[usenames,dvipsnames]{color}
\usepackage{comment}
\usepackage{ifpdf}
\usepackage{setspace}
\usepackage[utf8]{inputenc}
\usepackage[english]{babel}
\usepackage{graphicx}  
\usepackage{breqn}
\usepackage{mathrsfs}
\usepackage[left=.8in,right=.8in,top=.8in,bottom=.8in]{geometry}  
\begin{document}

\title{Towards Black Hole Entropy in Shape Dynamics}   
\author{Gabriel Herczeg$^\star$ and Vasudev Shyam$^\dagger$}
\affil{$^\star$Department of Physics, University of California, Davis}
\affil{$^\dagger$Perimeter Institute for Theoretical Physics}

\date{\today}      
\maketitle

\begin{abstract}
\noindent Shape dynamics is a classical theory of gravity which agrees with general relativity in many important cases, but possesses different gauge symmetries and constraints. Rather than spacetime diffeomorphism invariance, shape dynamics takes spatial diffeomorphism invariance and spatial Weyl invariance as the fundamental gauge symmetries associated with the gravitational field. Since the area of the event horizon of a black hole transforms under a generic spatial Weyl transformation, there has been some doubt that one can speak sensibly about the thermodynamics of black holes in shape dynamics. The purpose of this paper is to show that by treating the event horizon of a black hole as an interior boundary, one can recover familiar notions of black hole thermodynamics in shape dynamics and define a gauge invariant entropy that agrees with general relativity.
\end{abstract}

\section{Introduction}
\subsection{General Background}
Shape dynamics is a classical, Hamiltonian theory that models gravitation as the time evolution of the three-dimensional conformal geometry of space. While it is closely related to the canonical formulation of general relativity due to Arnowitt, Deser and Misner (ADM) \cite{ADM}, and even reproduces this description of the gravitational field in many important instances, it does not assume the existence of a smooth spacetime geometry. In fact, it has recently been shown that the simplest asymptotically flat black hole solutions admitted by shape dynamics disagree with general relativity at and within their event horizons, and that a smooth spacetime geometry fails to emerge on the horizon. Moreover, the areas of the horizons in the shape dynamic black hole solutions are not invariant under spatial Weyl transformations. The fact that these areas are not gauge-invariant quanitities has cast considerable doubt on the prospect of recovering the famous result that the entropy of a black hole is equal to one quarter of the horizon area. The  purpose of this paper is to show that by carefully treating the horizon as an interior spatial boundary and analyzing the counterterms needed to obtain well-defined equations of motion in the sense first described by Regge and Teitelboim \cite{Regge}, one can indeed recover the result that $S=A/4$ for a shape dynamic black hole. 

This is an important result for shape dynamics in light of the fact that the proportionality of black hole entropy and horizon area has been derived independently by state-counting arguments in a diverse array of approaches to quantum gravity each of which count very different states to arrive at the same result. The fact that these very different approaches universally yield the same result is something of an interesting problem in and of itself, and it has been argued by Carlip \cite{Carlip, universality} and others that this ``problem of universality" may arise as a result of a near-horizon conformal symmetry in the Einstein-Hilbert action. At any rate, the ubiquity of the thermodynamic properties of black holes has been so thoroughly demonstrated at the theoretical level, that while we have yet to detect Hawking radiation by direct observation, one would rightly be deeply skeptical of any model of a black hole that does not admit a thermodynamic interpretation.

\subsection{Shape Dynamics}
The most straightforward way to construct shape dynamics is to make use of a tool called the ``linking theory" that serves as a theoretical bridge between the ADM formulation of general relativity and shape dynamics. One starts with the ADM phase space, and then extends this phase space by a new pair of conjugate variables $(\phi,\pi_{\phi})$. In more concrete terms, we begin with ADM action, and perform the canonical embedding into the phase space of the linking theory defined by
\begin{align}
g_{ab} \to t_{\phi}(g_{ab}) := e^{4\phi}g_{ab} \label{embed1} \\
\pi^{ab} \to t_{\phi}(\pi^{ab}) := e^{-4\phi}\pi^{ab} \label{embed2}
\end{align}

\noindent where $g_{ab}$ is the spatial metric induced on a constant-$t$ hypersurface, and $\pi^{ab}$ is its conjugate momentum. This procedure yields the action of the linking theory, which now posseses not only the standard ADM constraints\footnote{The quantity $G_{abcd}$ appearing in the scalar constraint $\mathcal{S}(x)$ is the DeWitt supermetric defined by: \\ $G_{abcd}:= \frac{1}{2}(g_{ac}g_{bd} + g_{ad}g_{bc})-g_{ab}g_{cd}.$ \\}

\begin{eqnarray}\label{ADM const.}
\mathcal{S}(x) &=& \frac{G_{abcd}\pi^{ab}\pi^{cd}}{\sqrt{g}} - R\sqrt{g} \\
\mathcal{H}_a(x) &=& \pi^a_{b;a}
\end{eqnarray}

\noindent which generate on-shell refoliations and spatial diffeomorphisms, but also a new ``conformal constraint"

\begin{equation}\label{conf. const.}
D(x) = g_{ab}\pi^{ab}
\end{equation}

\noindent which generates spatial Weyl transformations. In order to obtain shape dynamics, one then imposes the gauge fixing condition $\pi_{\phi} = 0$, which reproduces the original phase space $(g_{ab},\pi^{ab})$. The resulting theory depends on the field $\phi$, but $\phi$ is no longer a dynamical variable. One can show that as a result of performing this phase space reduction, the scalar constraint $\mathcal{S}(x)$ is no longer first class with respect to the diffeomorphism constraint $H_a(x)$ and the conformal constraint $D(x)$ which are the only remaining first class constraints in shape dynamics. The result is a theory which, loosely speaking, is defined on a fixed maximal foliation, where the conformal constraint $D(x) = 0$ can be identified with the maximal slicing condition commonly used for finding valid initial data in the ADM formulation of general relativity. What distinguishes shape dynamics from ADM general relativity in maximal slicing is that the role of the gauge fixing conditions and constraints have been reversed---$D(x)$ is now viewed as a first class constraint generating spatial Weyl transformations rather than as a gauge-fixing condition on the ADM phase space. For a more details on the construction and historical development of shape dynamics see \cite{GGK, poincare, Linking, FlavioTutor, superspace}. It is worth noting that the form of the constraints and equations of motion for shape dynamics differ somewhat depending on the spatial topology. Throughout this paper we assume asymptotically flat boundary conditions on a spatially non-compact topology as in \cite{poincare}. For details on the spatially compact case see e.g. \cite{GGK}.

\subsection{Shape Dynamic Black Holes}
It has recently been shown \cite{Birkhoff,Kerr} that the simplest stationary, asymptotically flat black hole solutions of the shape dynamics equations of motion differ from their general relativistic counterparts at and within their event horizons. The reason for this disagreement is actually quite simple. One can see straightforwardly from the linking theory that shape dynamics and general relativity can generically be viewed as gauge fixings of each other: by imposing maximal slicing on a solution of the ADM equations of motion one generically obtains a local solution of the shape dynamics equations of motion, while by imposing York scaling\footnote{By York scaling we mean that the conformal factor $\phi$ obeys the Lichnerowicz-York equation:\\ $\nabla^2\Omega + \frac{R}{8}\Omega - \frac{1}{8}\frac{\pi^{ij}\pi_{ij}}{|g|}\Omega^{-7} = 0$ where $\Omega = e^{\phi}$.} for the conformal factor $\phi$ on a solution of the shape dynamics equations of motion one generically obtains a local solution of the ADM equations of motion. We use the qualifier ``generically" because this duality between general relativity and shape dynamics can break down if there is a surface on which the lapse vanishes, such as the bifurcation surface of a stationary black hole. Since the determinant of the spacetime metric is of the form $\sqrt{g^{(4)}} = N\sqrt{g}$, one can immediately see that if the lapse goes to zero on some lower dimensional subset, then the spacetime metric and the spatial metric cannot be simultaneously invertible there---one must choose between a smooth spacetime geometry and smooth spatial conformal geometry, and each choice dictates a distinct smooth continuation across the surface of vanishing lapse. The fact that the duality between shape dynamics and general relativity and the emergence of local Lorentz invariance in shape dynamics require a non-vanishing lapse has been explored in detail in \cite{Lorentz}.    

It was shown by Gomes in \cite{Birkhoff} that shape dynamics admits a Birkhoff theorem. The unique spherically symmetric, asymptotically flat solution to the shape dynamics equations of motion was shown to be described globally by the Schwarzschild line element in isotropic coordinates:

\begin{equation}\label{isotropic}
ds^2= -\left(\frac{1-\frac{m}{2r}}{1+\frac{m}{2r}}\right)^2dt^2+ \left(1+\frac{m}{2r}\right)^4\left(dr^2+r^2(d\theta^2+\sin^2\theta d\phi^2)\right).
\end{equation}

\noindent We stress once again that this line element is a solution of the Einstein field equations \emph{only} for $r \neq m/2$, where $m/2$ is the location of the event horizon in isotropic coordinates, whereas this solution is valid \emph{everywhere} from the point of view of shape dynamics. It is straightforward to check that this solution possesses an inversion symmetry about the horizon, which can be seen by noting that the line element is invariant under the transformation  $r\to m^2/(4r)$. An immediate consequence of this inversion symmetry is that the ``interior" of this solution is a time-reversed copy of the exterior, giving the solution the character of \emph{wormhole}. An obvious consequence of the wormhole character of the solution is that it possesses no central curvature singularity. In fact, this solution is completely free of physical singularities, which from the point of view of shape dynamics means that there are no singularities in the spatial conformal structure. This can be seen immediately by noting that the solution is spatially conformally flat, which means that the Cotton tensor\footnote{The Cotton tensor is defined by $\mathscr{C}_{abc} := \nabla_c\left(R_{ab} - \frac{1}{4}Rg_{ab}\right) - \nabla_b\left(R_{ac} - \frac{1}{4}Rg_{ac}\right)$, and it is a well known fact that it possesses all of the local information on the conformal structure of a three dimensional Reimannian manifold just as the Weyl tensor contains all of the local information on the conformal structure of higher dimensional manifolds \cite{Gourgoulhon}.} vanishes identically.

It was shown in \cite{Kerr} that shape dynamics admits a rotating black hole solution with many of the same interesting properties found in the spherically symmetric case. The rotating solution also possesses an inversion symmetry about the horizon manifesting its wormhole character, and it was argued in \cite{Kerr} that this solution is also free of physical singularities by analzying the behavior of a particular scalar density constructed from the Cotton tensor. The reconstructed line element associated with the rotating solution was found to be:

\begin{equation}\label{Kerr PS} 
ds^2 = -\lambda^{-1}(dt - \omega d\phi)^2 + \lambda [m^2e^{2\gamma}(d\mu^2 + d\theta^2) + s^2d\phi^2]
\end{equation}
where 
\begin{eqnarray}\label{Kerr functions PS} 
s \hspace{5pt} &=& mp\sinh\mu\sin\theta \nonumber \\
e^{2\gamma} &=& p^2\cosh^2\mu  + q^2\cos^2\theta - 1 \\
\omega \hspace{5pt} &=& e^{-2\gamma}\left[2mq\sin^2\theta(p\cosh\mu + 1)\right] \nonumber \\
\lambda \hspace{5pt} &=& e^{-2\gamma}\left[(p\cosh\mu + 1)^2 + q^2\cos^2\theta \right]. \nonumber
\end{eqnarray} 

\noindent Here $p^2 + q^2 = 1$, and the solution becomes spherically symmetric in the limit $p \to 1$, $q \to 0$. In the prolate spheroidal coordinates $(\mu,\theta)$, the inversion symmetry about the horizon can be seen by noting that the line element is invariant under $\mu \to -\mu$. It is interesting to note that in addition to avoiding the ring-like curvature singularity of the Kerr spacetime, this solution also avoids the closed timelike curves and inner horizons of its sister solution, giving it a generally much simpler structure. Singularity avoidance theorems for ADM in maximal slicing make it likely that solutions of shape dynamics very generally avoid physical singularities, and the fundamentally spatial perspective of the theory makes it likely that solutions of shape dynamics generally do not develop closed timelike curves.\footnote{In fact, the global hyperbolicity of the regions admitting spacetime reconstruction virtually guarantees this.} Detailed investigations of singularity avoidance and chronology protection in shape dynamics are left for future work.

\section{Boundary Hamiltonian, Euclidean Gravity and Black Hole Entropy}
It was first pointed out by Regge and Teitelboim \cite{Regge} that in order to properly formulate canonical general relativity with spatially non-compact boundary conditions, one must add a boundary term $H_B$ to the pure constraint ADM Hamiltonian $H_0$ so that the total Hamiltonian is given by the sum 

\begin{equation}
H = H_0 + H_B.
\end{equation}

\noindent If one neglects the boundary term $H_B$ no well-defined equations of motion are generated by $H_0$---only the total Hamiltonian $H  = H_0 + H_B$ yields well-defined equations of motion. This is essentially due to the fact that when calculating Poisson brackets on a manifold with boundary, one cannot discard the boundary terms that arise due to integration by parts. Rather, these terms must be canceled with appropriate counterterms so that Hamilton's equations can be recovered. Moreover, given mild restrictions on the fall-off of the surface deformations, the boundary term $H_B$ can be identified with the total energy of the system. Let $\Sigma$ be a constant-$t$ hypersurface with boundary $\partial\Sigma$. Including all boundary terms, the ADM action can be written as

\begin{equation}\label{ADM action}
\mathcal{I}_{\mbox{\tiny ADM}} = \int dt\bigg[\frac{1}{16\pi} \int_{\Sigma} d^3x\sqrt{g} \left(\dot{g}_{ab}\pi^{ab} -
  N\mathcal{S}(x) - \xi^a\mathcal{H}_a(x)\right)  -
\frac{1}{8\pi}\int_{\partial\Sigma} d^2x\sqrt{\sigma}(NK+r^aN_{,a}-r_a\xi^b\pi^a_b) \bigg] 
\end{equation}

\noindent where $\sigma_{ab}$ is the metric induced on $\partial\Sigma$ by $g_{ab}$, $r^a$ is the outward pointing normal vector of  $\partial\Sigma$, $K$ is the trace of the extrinsic curvature of $\partial\Sigma$ embedded in $\Sigma$, and $N$ and $\xi^a$ are the lapse function and shift vector. Note that in \eqref{ADM action} we have written the total action in the form where the Legendre transformation is most easily carried out so that one can essentially read the Hamiltonian directly off of this equation.

The connection between gravity and thermodynamics was first proposed by Bekenstein in 1972 \cite{Bekenstein} and confirmed by Hawking in 1974 through the discovery of thermal black hole radiation (Hawking radiation) \cite{Hawking Radiation}. In 1976, Gibbons and Hawking introduced the  Euclidean canonical formalism in \cite{Hawk}. In this work, it was shown that the path integral associated with the Euclideanized gravitational action has the exact form of a canonical partition function at inverse temperature $\beta$ where the path integral is taken over all field configurations that are periodic with period $i\beta$. From the lowest order contributions to the canonical partition function, one can then obtain all of the thermodynamic quantities associated with the horizons of classical solutions to the Einstein field equations, including the entropy.  

Since these seminal works, many other approaches to black hole thermodynamics have been developed.  The approach that will be most useful for us was developed by Padmanabhan \cite{HoloPad} and others, and identifies the boundary action (or Hamiltonian) evaluated on the horizon of a stationary black hole with the entropy (or the temperature times the entropy) of the black hole. In this approach, one considers the fact that horizons generically appear for some families of observers (such as accelerated observers in flat spacetime, and Eulerian observers in the exterior of the Schwarzschild spacetime) but not for others (such as inertial observers in flat spacetime or free-falling observers in the Schwarzschild spacetime). One then argues that any family of observers should be able to use an action principle that makes use only of the information available to those observers. Since observers who see a horizon cannot recieve information from the other side of the horizon, the action integral for observers who see horizons should only depend on the portion of the spacetime outside the horizon. Since there is a ``tracing out" of information about what is on the other side of the horizon when one uses such a family of observers, it is natural to identify the boundary term in the action coming from the horizon with an entropy, and one indeed finds that for a stationary horizon with $N = 0$, $N_{,a} =  \kappa r_a$ and $\pi^{ab} = 0$ where $\kappa$ is the surface gravity of the horizon, the boundary action equals one quarter of the horizon area.  

Now we would like to show that the same argument can be used to relate the horizon contribution to the shape dynamics Hamiltonian with the entropy of a shape dynamic black hole. The boundary Hamiltonian for shape dynamics was derived in \cite{poincare} and its variation was found to be 

\begin{multline}\label{Boundary Variation 1}\delta t_\phi H_B(N,\xi)=2\int_{\partial\Sigma} d^2 y\Big( \xi^ar^b\left(\pi^{cd}\left((5g_{ac}g_{bd}-g_{ab}g_{cd})\delta\phi+(g_{bd}\delta g_{ca}-\frac{1}{2}g_{ab}\delta g_{cd})\right)+g_{ac}g_{bd}\delta\pi^{cd}\right)\\
+\int_{\partial\Sigma} d^2 y\sqrt h e^{2\phi}\Big(8\left( N\delta\phi^{,a}-N^{,a}\delta\phi\right)r_ a+ \left( N\delta g_{ab;d}+(6\phi_{,d} N-N_{,d})\delta g_{ab}\right)(g^{de}g^{ab}-g^{da}g^{be}) r_e \Big).
\end{multline}

\noindent  Eulerian observers\footnote{ An Eulerian observer, also known as a hypersurface orthogonal observer or zero angular momentum observer (ZAMO) is an observer whose four-velocity $u^{\mu}$ is perpendicular to $\Sigma_t$, i.e $u^{\mu}\epsilon_{\mu\nu\rho} = 0$, where $\epsilon_{\mu\nu\rho}$ is the volume form on $\Sigma_t$.} are natural observers to choose in shape dynamics because they are at rest with respect to the spatial geometry. These observers see a horizon in stationary black hole solutions, so it is appropriate to include a boundary contribution to the Hamiltonian from the horizon. For Eulerian observers, one can use the same boundary conditions used in general relativity:  $N = 0$, $N_{,a} =  \kappa n_a$ and $\pi^{ab} = 0$. When these boundary conditions are imposed, the total variation of the boundary Hamiltonian becomes:

\begin{gather}
\delta t_{\phi}H_B(N,\xi) = -\frac{\kappa}{16\pi}\int_{\partial\Sigma}d^2
y\sqrt{\sigma}e^{2\phi}\left( 8\delta\phi +
  r_dr_e\delta g_{ab}(g^{de}g^{ab}- g^{da}g^{be})  \right) \nonumber \\
=  -\frac{\kappa}{16\pi}\int_{\partial\Sigma}d^2
y\sqrt{\sigma}e^{2\phi}\left( 8\delta\phi + (g^{ab} - r^ar^b)\delta
  g_{ab} \right) \label{boundary variation 2} \\
= -\frac{\kappa}{16\pi}\int_{\partial\Sigma}d^2
y\sqrt{\sigma}e^{2\phi}\left(8\delta\phi + \sigma^{ab}\delta
  g_{ab} \right). \nonumber \\ \nonumber
\end{gather}

\noindent On the other hand, 

\begin{eqnarray}\label{detVar}
\delta\sqrt{\sigma} &=& \frac{1}{2}\sqrt{\sigma}\sigma^{ab}\delta\sigma_{ab} =
\frac{1}{2}\sqrt{\sigma}\sigma^{ab}\delta\left( g_{ab} - r_ar_b 
\right) \nonumber \\
&=& \frac{1}{2}\sqrt{\sigma}\sigma^{ab}\left(\delta g_{ab} -
  r_ar^c\delta g_{bc} - r_br^c\delta g_{ac} \right) \nonumber \\
&=&  \frac{1}{2}\sqrt{\sigma}\sigma^{ab}\delta g_{ab}.
\end{eqnarray}

\noindent Equation \eqref{detVar} shows that the factor of $8$ appearing in the first term of the last line of \eqref{boundary variation 2} should actually be a $4$ in order for  \eqref{boundary variation 2} to be a total variation. It is absolutely essential that  \eqref{boundary variation 2} be a total variation in order for the boundary conditions we imposed to be consistent, since this term is \emph{by definition} the variation of the boundary contribution to the Hamiltonian. This suggests that we need to impose additional boundary conditions on the conformal factor $\phi$ at the horizon. An obvious choice is $\phi = 0$, $\delta \phi = 0$ since then we would fully break Weyl invariance on the
horizon, and our horizon Hamiltonian would then be simply

\begin{equation}\label{TS}
H_{hor}(N) = \frac{\kappa}{8\pi}\int_{\partial\Sigma}d^2y\sqrt{\sigma} = \frac{\kappa}{2\pi}\cdot\frac{A}{4}
\end{equation}

\noindent where $A$ is the same horizon area we would have found in general relativity. Now, if we identify $T =  \frac{\kappa}{2\pi}$, $S = \frac{A}{4}$, we find complete agreement with the horizon thermodynamics of general relativity. 

One might worry, however,
that this condition is too restrictive. We could consider restricting
$\delta\phi$ by requiring that it depend on $\delta \sigma_{ab}$. For
example, if we demand that $\delta\phi = \lambda\sigma^{ab}\delta\sigma_{ab}$ for some constant $\lambda$. However, this
approach would require the conformal factor to transform by a
$\lambda$-dependent rescaling on the horizon, which can be shown to be inconsistent with our boundary conditions at the level of the
constraints. The source of this inconsistency comes from the fact that
the bracket of the horizon Hamiltonian with the conformal constraint
can vanish only if $\langle\rho\rangle_{\sigma} :=
\frac{\int_{\partial\Sigma}d^2y\sqrt{\sigma}\rho}{\int_{\partial\Sigma}d^2y\sqrt{\sigma}}
= 0$, which implies that the spatial Weyl transformations which are pure
gauge in the presence of a horizon with the boundary conditions we
have specified are exactly those that preserve the area of the
horizon. 

In anticipation of this restriction on the Weyl invariance of shape
dynamics in the presence of a horizon, we can obtain the same outcome
as if we had set $\delta\phi = 0$ without having to make quite such a
rigid restriction. Suppose we require that

\begin{equation}
\int_{\partial\Sigma}d^2y\sqrt{\sigma}\delta\left(e^{2\phi}\right)
= 0.
\end{equation}

\noindent If we further demand that $\phi = 0$ lies in our solution space on the
horizon, so that we may recover a non-trivial intersection with GR,
then it is clear that what we are really demanding is not that $\phi$ be fixed on the horizon, but rather that the allowed
transformations of $\phi$ are those that preserve the horizon area. With these additional restrictions in place, we can integrate
\eqref{boundary variation 2} to obtain

\begin{equation}
H_{hor}(N) = \frac{\kappa}{8\pi}\int_{\partial\Sigma}d^2y\sqrt{\sigma}e^{2\hat{\phi}}
\end{equation}

\noindent where $\hat{\phi} :=
\frac{1}{2}\ln\langle e^{2\phi}\rangle_{\sigma}$, $\phi$ is an
arbitrary conformal factor satisfying the usual asymptotically flat
boundary conditions at spatial infinity, and $\langle \cdot \rangle_{\sigma}$ denotes the average over the horizon with respect to the induced metric $\sigma$. Due to the area-preserving
nature of the allowed Weyl transformations, the horizon Hamiltonian
can be written with or without the smearing $e^{2\hat{\phi}}$:

\begin{equation}
H_{hor}(N) = \frac{\kappa}{8\pi}\int_{\partial\Sigma}d^2y\sqrt{\sigma}
\end{equation}

\noindent which is identical the corresponding result for general
relativity. However, here we had to impose boundary conditions in two
stages. First, we imposed boundary conditions on the momentum, and on
the lapse and its normal derivative, in order to capture the
essential features of a horizon. Next, we were forced to impose
boundary conditions on the conformal factor as a consistency condition for
the functional integrability of the total boundary variation
\eqref{Boundary Variation 1}. 

Next, we will show that the horizon Hamiltonian we have
derived puts restrictions on the smearing $\rho$ of the conformal
constraint $D(\rho)$, by calculating the bracket $\{H_{hor}(N), D(\rho)\}$
and observing that it only vanishes for a certain class of smearings
$\rho$. We find (see appendix) that

\begin{gather}\label{total bracket}
\{H_{hor}(N), D(\rho)\} = -\frac{3}{16\pi}\int_{\Sigma}d^3x
\frac{\rho N}{\sqrt{g}}e^{-6\hat{\phi}} G_{abcd}\pi^{ab}\pi^{cd} =
-\frac{3}{2}H_{hor}(\rho N).
\end{gather}

Since the horizon Hamiltonian is weakly non-vanishing, \eqref{total bracket} tells us that the bracket between the horizon Hamiltonian and the conformal constraint cannot weakly vanish for arbitrary smearings $\rho$. It is possible to derive an elliptic differntial equation for $\rho$ so that \eqref{total bracket} vanishes, but this characterization of the allowed spatial Weyl transformations is not particularly illuminating and technically difficult to analyze. However, it is possible to ``softly" modify the constraints of theory in such a way that we can clearly show that in the presence of a horizon satsfying the boundary conditions above, the spatial Weyl transformations that are pure gauge are precisely those that preserve the horizon area. This will be the subject of the following section.

\section{Modified Constraints and Area-Preserving Weyl Transformations}
Another way to obtain a gauge-invariant horizon entropy for shape dynamics is to softly modify the the conformal constraint in such a way that the modification repects the boundary conditions and leaves the horizon area invariant. The simplest realization of these conditions is in writing the conformal constraint in the following manner:
\begin{equation}\label{modConstr}D(\rho)=\int_{\Sigma}\textrm{d}^{3}x \rho g_{ab}\pi^{ab} -\int_{B}\textrm{d}^{2}y\rho \langle g_{ab}\pi^{ab} \rangle_{\sigma}.\end{equation}
Here we have just rewritten the usual conformal constraint with the addition of a new term which is identically zero on the interior boundary due to the boundary conditions used to characterize the horizon, i.e. given that $\pi^{ab}|_{B}=0$ the term within the angular brackets is strongly equal to 0. Note that $\frac{\delta D(\rho)}{\delta \rho}=0=g_{ab}\pi^{ab}$ keeping in mind the boundary condtitions imposed, so we are ensured that the constraint is not altered in any violent manner. It is convenient to introduce the weakly non degenerate symplectic structure on the shape dynamics phase space to ease our computations. The symplectic structure for shape dynamics is defined by the two-form
\begin{equation}\Omega_{SD}=\int_{\Sigma} \delta \pi^{ab}\wedge \delta g_{ab}.\end{equation}
The Hamiltonian vector field $X_{f}$ correponding to an arbitrary phase space function $f[g_{ab},\pi^{ab}]$ is defined through the equation
$$\Omega_{SD}(X_{f})=\delta f.$$
The integral curves of the Hamiltonian vector field correpond to its flow on phase space, and the vector field is identified with the generator of infinitesimal transformations along said flow. When this vector field corresponds to a first class constraint, it generates gauge transformations along the orbits of this constraint on phase space. The Hamiltonian vector field for
the conformal transformations is thus given by 
\begin{equation}X_{D(\rho)}=(\rho-\langle \rho\rangle_{\sigma})g_{ab}\frac{\delta}{\delta g_{ab}}+(\rho-\langle \rho\rangle_{\sigma})\pi^{ab}\frac{\delta}{\delta \pi^{ab}},\end{equation}
which satisifies the equation $$\Omega_{SD}(X_{D(\rho)})=\delta D(\rho).$$
Thus, the object given by \eqref{modConstr} generates infinitesimal spatial Weyl transformations on the shape dynamics phase space. A simple calculation yields
\begin{equation}\label{apres}X_{D(\rho)}A=0.\end{equation}
Where $A$ is the area of the horizon. This unambiguously shows that if the right form of the conformal constraint is chosen keeping in mind the boundary conditions, the seemingly ``soft" modification of the constraint produces the correct flow on phase space. Given that the Weyl transformations generated by the constraint given by \eqref{modConstr} preserves the horizon area, in terms of large spatial Weyl transformations, \eqref{apres} is equivalent to the statement
$$\int e^{2\phi}\sqrt{\sigma} \textrm{d}^{2}y=\int\sqrt{\sigma} \textrm{d}^{2}y=A.$$
This implies for the variation of the conformal factor at the boundary that:
\begin{equation}\int_{B}\delta \phi\sqrt{\sigma}=0.\end{equation}
Going back to \eqref{boundary variation 2}, we find that now, 
$$\delta t_{\phi}H_B(N,\xi)=-\frac{\kappa}{8\pi}\int_{\partial \Sigma}\textrm{d}^{2}y \sqrt{\sigma} \sigma^{ab}\delta g_{ab},$$
as in general relativity, and from this, we can obtain the surface Hamiltonian for the horizon 
$H_{hor}=TS=\frac{\kappa A}{8\pi}.$
Identifying the Hawking--Unruh temperature $T=\frac{\kappa}{2\pi},$ the entropy is:
\begin{equation}S=\frac{A}{4},\end{equation}
which is completely identical to the area--entropy law of General Relativity, just as we derived by more direct but less elegent means in the preceding section.

\section{Discussion}
We have shown that by correctly identifying the spatial Weyl transformations that remain pure gauge in the presence of a horizon which acts as an interior spatial boundary, it is possible to identify the gauge-invariant quantity that we have suggestively labeled $S$ and which we believe plays the role of a thermodynamic entropy associated with a shape dynamic black hole. Furthermore, if we assume that that this quantity really is the entropy associated with a shape dynamic black hole, then we can reproduce the standard area-entropy relation of general relativity and various approaches to quantum gravity. It is worth stressing that while our $S$ seems to have many features that suggest it can be interpreted as an entropy (it arises from ``tracing out" information about the interior that exterior observers do not have access to, it agrees with standard results from black hole thermodynamics, it is gauge-invariant, etc.) the authors must admit that until it can be shown that shape dynamic black holes emit Hawking radiation and that the temperture and entropy agree with the $T$ and $S$ we have identified, there is insufficient evidence to conclude that these quantities really play the role of a physical temperature and entropy. 


It was conjectured in \cite{Kerr} that since the equivalence principle is an emergent property of shape dynamics, rather than an axiom as in general relativity, and that this property fails to emerge precisely on the event horizon of a shape dynamic black hole, that a quantum theory of gravity based on a canonical quantization of shape dynamics might present a possible resolution to the firewall paradox introduced by Almheiri et. al. \cite{AMPS}. In order to address this question in a more systematic manner, it is once again essential to understand the thermodynamic properties of shape dynamic black holes including the radiative properties of quantum fields in a shape dynamic black hole background (i.e. Hawking radiation).

The next step in the investigation of the thermodynamic properties of shape dynamic black holes is to determine whether shape dynamic black holes produce Hawking radiation. A preliminary analysis of the features of shape dynamic black holes and the mechanism by which black holes produce Hawking radiation in general relativity makes it seem very likely that shape dynamic black holes produce thermal Hawking radiation in essentially the same way as their general relativisic counterparts. Ordinary Hawking radiation can be realized by coupling a scalar field to the Einstein-Hilbert action and analyzing the resulting covariant wave equation on the black hole background. One then identifies the zeroes of the Regge-Wheeler potential where the covariant wave equation reduces to the usual Klein-Gordon equation in flat spacetime, and expands the solutions in Fourier modes. Since the zeroes of the Regge-Wheeler potential occur at spatial infinity and on the event horizon of a Schwarzschild black hole, and since the exterior region of a Schwarzschild black hole is isomorphic to the exterior of the spherically symmetric black hole solution derived in \cite{Birkhoff}, it seems very likely that standard Hawking radiation can be recovered for shape dynamic black holes. We leave a more careful and systematic treatment of this problem for future work.

\subsection*{Acknowledgments}
We would like to thank Henrique Gomes for helpful conversations. We are also grateful to Sean Gryb for lively discussions and for his generous hospitality during the final stages of this work, and to Flavio Mercati for helpful suggestions. G.H. would like to thank Steve Carlip for his valuable insights and suggestions. G.H was supported in part by FQXi minigrant MGA-1404 courtesy of the Foundational Questions Institute. V.S. was supported in part by Perimeter Institute for Theoretical Physics. Research at Perimeter Institute is supported by the Government of Canada through Industry Canada and by the Province of Ontario through the Ministry of Research and Innovation.

\subsection*{Appendix}
 In order to calculate the bracket $\{H_{hor}(N), D(\rho)\}$, we will first need to
transform $H_{hor}(N)$ back into a volume integral:

\begin{gather}
H_{hor}(N)
=
\int_{\partial\Sigma}d^2y\sqrt{\sigma}\frac{\kappa}{8\pi}e^{2\hat{\phi}}
\nonumber\\ 
=
\frac{1}{8\pi}\int_{\partial\Sigma}d^2y\sqrt{\sigma}n^a\left(\nabla_aNe^{2\hat{\phi}}\right) \label{bulk
H hor}\\
=
\frac{1}{8\pi}\int_{\Sigma}d^3x\sqrt{g}\hspace{.1cm}\nabla^a\left(\nabla_aNe^{2\hat{\phi}}\right) \nonumber
\end{gather}

\noindent Since the covariant derivative $\nabla_a$ depends on the metric, it is
not immediately clear how to treat it under the Poisson
bracket. Usually, this can be taken care of using repeated integration
by parts and making use of compactness or boundary conditions. Here
things are complicated by the presence of the extra boundary at the
horizon, so this is not the most convenient way to compute the
bracket. Rather, we make use of the lapse-fixing equation obtained from
phase-space reduction:

\begin{equation}\label{LFE}
\sqrt{g}\nabla^a\left(\nabla_aNe^{2\hat{\phi}}\right) = e^{-6\hat{\phi}}\frac{N}{\sqrt{g}}G_{abcd}\pi^{ab}\pi^{cd}.
\end{equation}

\noindent Inserting \eqref{LFE} into \eqref{bulk H hor} and acting with the conformal
constraint $D(\rho) := \int_{\Sigma}d^3x\rho g_{ab}\pi^{ab}$, we obtain:

\begin{gather}
\{H_{hor}(N), D(\rho)\} = \bigg\{\left(
\frac{1}{8\pi}\int_{\Sigma}d^3x\hspace{.1cm}e^{-6\hat{\phi}}\frac{N}{\sqrt{g}}G_{abcd}\pi^{ab}\pi^{cd}
\right),\left( \int_{\Sigma}d^3x^{\prime}\rho
  g_{ef}\pi^{ef}\right)\bigg\} \nonumber \\
=
\frac{1}{8\pi}\int_{\Sigma}d^3x\int_{\Sigma}d^3x^{\prime}
\bigg[
\left(e^{-6\hat{\phi}}N\pi^{ab}\pi^{cd}\right)(x)\left(\rho
g_{ef}\right)(x^{\prime})
\bigg\{\left(\frac{G_{abcd}}{\sqrt{g}}\right)(x)
,\pi^{ef}(x^{\prime})\bigg\} \label{bracket1}\\
+ \left(  e^{-6\hat{\phi}}\frac{N}{\sqrt{g}}G_{abcd}\right)(x)\left(\rho\pi^{ef}\right)(x^{\prime})
\bigg\{\left(\pi^{ab}\pi^{cd}\right)(x) ,g_{ef}(x^{\prime})\bigg\}
\bigg]. \nonumber
\end{gather}

\noindent By direct computation, we find

\begin{gather}
\pi^{ab}\pi^{cd}g_{ef}\bigg\{\left(\frac{G_{abcd}}{\sqrt{g}}\right)(x)
,\pi^{ef}(x^{\prime})\bigg\} =
-\frac{G_{abcd}}{2\sqrt{g}}\pi^{ab}\pi^{cd}\delta(x-x^{\prime}) \\
\frac{G_{abcd}}{\sqrt{g}}\pi^{ef}\bigg\{\left(\pi^{ab}\pi^{cd}\right)(x)
,g_{ef}(x^{\prime})\bigg\} = -\frac{2G_{abcd}}{\sqrt{g}}\pi^{ab}\pi^{cd}\delta(x-x^{\prime})
\end{gather}

\noindent so that we are finally left with 

\begin{gather}\label{last bracket}
\{H_{hor}(N), D(\rho)\} = -\frac{3}{16\pi}\int_{\Sigma}d^3x
\frac{\rho N}{\sqrt{g}}e^{-6\hat{\phi}} G_{abcd}\pi^{ab}\pi^{cd} =
-\frac{3}{2}H_{hor}(\rho N).
\end{gather}

\end{document}